\documentclass[12pt]{iopart}
\usepackage{iopams}

\newtheorem{proposition}{Proposition}

\def\di{\mathop{\textrm{d}}\nolimits}

\begin{document}

\title[Comment]
%{Comment on "Thermodynamic scheme of inhomogeneous perfect fluid
%mixtures"}
%
{COMMENT:\\
\vspace{2mm}
On the thermodynamics of inhomogeneous perfect fluid mixtures}
\author{Bartolom\'e Coll$^1$\
and Joan Josep Ferrando$^2$}
%\footnote[3]{To
%whom correspondence should be addressed (joan.ferrando@uv.es)}}

\address{$^1$\ Syst\`emes de r\'ef\'erence relativistes, SYRTE,
Obsevatoire de Paris-CNRS, \\75005 Paris, France.}

\address{$^2$\ Departament d'Astronomia i Astrof\'{\i}sica, Universitat
de Val\`encia,\\46100 Burjassot, Val\`encia, Spain.}
\ead{bartolome.coll@obspm.fr; joan.ferrando@uv.es}

\begin{abstract}
It is shown that inhomogeneous Szekeres and Stephani universes
exist corresponding to non-dissipative binary mixtures of perfect
fluids in local thermal equilibrium. This result contradicts a
recent statement by Z\'arate and Quevedo (2004 Class. Quantum
Grav. {\bf 21} 197, {\it Preprint} gr-qc/0310087), which affirms
that the only Szekeres and Stephani universes compatible with
these fluids are the homogeneous Friedmann-Robertson-Walker
models. Thus, contrarily to their conclusion, their thermodynamic
scheme do not gives new indications of incompatibility  between
thermodynamics and relativity. Two of the points that have
generated this error are commented.
\end{abstract}

%Uncomment for PACS numbers title message
\pacs{04.20.-q, 04.40.Nr, 05.70.-a}
% Uncomment for Submitted to journal title message
%\submitto{\CQG}
% Comment out if separate title page not required
%\maketitle
\vspace{1cm}

\section{Thermodynamic scheme for a mixture of two perfect fluids}

In a recent paper,  Z\'arate and Quevedo \cite{zq} extend the
standard thermodynamic scheme, of local thermal equilibrium for a
simple perfect fluid, to the case of a mixture of two perfect
fluids.

They define their thermodynamic scheme by the following relations:

    \indent(a) conservation of the perfect fluid energy-momentum tensor:
\begin{equation}   \label{conT}
    \nabla \cdot T = 0  \, , \qquad  T = (\rho + p) u \otimes u + p g       \, ,
\end{equation}
where $\rho$ is the total energy density, $p$ the mixture
pressure and $u$ the 4-velocity of the total matter flow.\\
    \indent(b) conservation of the total current density:
\begin{equation}   \label{conN}
    \nabla \cdot (n u) = 0  \, ,
\end{equation}
where $n$ is the total particle number density.\\
    \indent(c) Gibbs thermodynamic relation:
\begin{equation} \label{gibbs}
    T \di s = \di(\rho/n) + p \di(1/n) - \mu \di c \, ,
\end{equation}
where $s$ is the entropy per particle, $T$ the temperature, $c$
the fractional concentration of one of the two perfect fluid
components and $\mu$ is the mixture chemical potential, i.e. the
difference between the chemical potentials of the components of
fractional concentration $c$ and $1 -c$ respectively.

Under these relations, and as Z\'arate and Quevedo emphasize, the
entropy production no longer vanishes but is a consequence of the
change in the fractional concentrations of the components.

In addition, this thermodynamic scheme for a mixture of two
perfect fluids recovers the standard scheme for a simple fluid
when $c=constant.$ In this sense their scheme is {\em more
general} than the standard one.

Nevertheless, its physical applicability risks to be much {\em
more restrictive} than that of a simple fluid: the physical
applicability of the non-dissipative (relativistic) evolution
hypothesis clearly diminishes generically when in addition $c\neq
constant$, i.e. the fluid is submitted to endothermic or
exothermic reactions. A more realistic model would be consisted in
a mixture of two perfect fluids generating a semi-perfect fluid,
that is to say, a dissipative Pascalian one, with eventually a
heat flux proportional to the gradient of the fractional
concentration.

Anyway, as no complete set of constraints is known on the space of
formal constitutive relations in continuous thermodynamics (i.e.
the  more or less general inequalities on the thermodynamic
variables and on some of their partial derivatives are
insufficient to separate physical from unphysical equations of
state), it makes sense, as Z\'arate and Quevedo emphasize, to
analyze the compatibility of their thermodynamic scheme with
Einstein equations.

\section{Compatibility of Z\'arate and Quevedo's thermodynamic
scheme with Einstein equations}

In the article \cite{zq} that motivates this comment, the authors
quote a work by us \cite{cfrainich} and they assert: "Coll and
Ferrando have shown that an exact solution [to Einstein equations]
admits a thermodynamic scheme provided the integrability
conditions of Gibbs equation are satisfied".

It is here pertinent to observe that, of course, this fact is the
starting point of the quoted paper, but that our main goal there
is {\em i)} to obtain these integrability conditions explicitly
and in terms of the sole hydrodynamic variables $(u,\rho,p),$ and
{\em ii)} to interpret them physically. Our corresponding result
is \cite{cfrainich} (see also \cite{cfERE1}): {\em i)} {\it a
divergence-free perfect fluid energy tensor evolves in local
thermal equilibrium if, and only if, the space-time function $\xi
\equiv \dot{\rho}/\dot{p}$ depends only on the variables $\rho$
and $p$: $\, \di \xi \wedge \di \rho \wedge \di p = 0 ;$  ii) then
$\xi$ is a state variable, $\xi = \xi(\rho,p),$ and represents the
square of the velocity of the sound}.

As, on one hand,  Einstein equations are biunivocally related to
the sole hydrodynamic variables $(u,\rho,p)$ and, on the other
hand, the condition $\, \di \xi \wedge \di \rho \wedge \di p = 0$
is not a consequence of them, it follows that: {\em there exist
perfect fluid solutions to Einstein equations that do not admit a
standard thermodynamic scheme, {\em i.e.} that cannot be
interpreted as evolutions, in local thermal equilibrium, of a
single fluid.}

What is the corresponding result for Z\'arate and Quevedo's
thermodynamic scheme for a non-dissipative mixture of two fluids?

As there exists only one scalar constraint for the compatibility
between Einstein equations and the relatively restrictive standard
thermodynamic scheme (namely, the one expressing that the quotient
$\,\dot \rho /\dot p\,$ is a function of state: $\,\dot \rho /\dot
p = \xi(\rho,p)\,)$, it seems that the corresponding result for
Z\'arate and Quevedo's scheme would be the absence of constraint,
because of its less restrictive character. But this appears as a
startling evaluation facing Z\'arate and Quevedo's main statement
in \cite{zq}. Let us see it in detail.

Suppose given, in a domain of the space-time, a solution to
Einstein equations for a perfect fluid, $(u,\rho,p)$ (in fact, a
solution to the divergence-free equations (\ref{conT}), the
argument that follows being also valid for test fluids).  Then,
$u,$  $\rho$ and  $p$ are known quantities in that domain and, in
particular, the conservation equation for the total current
density (\ref{conN}) becomes a linear and homogeneous scalar
equation in the scalar unknown $n.$ Pick in it a particular
solution $n;$ with these four known elements $(u,\rho,p,n)$ we can
evaluate the one-form $\omega$ defined by
\begin{equation} \label{omega}
    \omega \equiv \di(\rho/n) + p \di(1/n)
\end{equation}
Then, according to Gibbs equation (\ref{gibbs}) for the mixture,
one has to explore the existence of the four thermodynamic scalars
$(s, c, T, \mu)$ such that
\begin{equation} \label{omega2}
    \omega = T \di s + \mu \di c
\end{equation}
But this existence is exactly what  Pfaff decomposition theorem
locally guarantees in four dimensions for {\em any} one-form
$\,\omega\,, $  and {\em a fortiori} for the restricted one given
by (\ref{omega}); we have thus:

\begin{proposition}
Any perfect fluid solution to Einstein equations is compatible
with a Z\'arate and Quevedo's binary thermodynamic scheme.
\end{proposition}

How many Z\'arate and Quevedo's schemes may be associate to any
given perfect fluid solution to Einstein equations?

From the data $\,(u,\rho,p)\,$ obtained as a solution to equation
(\ref{conT}), the solution to the density conservation
(\ref{conN}) is obviously determined up to an $u$-invariant
function $\,f\,,$ $\, \dot{f}\equiv  u^\alpha\partial_\alpha f =
0\,,$ so that  $\,n = fn_o\,,$ $\,n_o\,$ being a particular
solution. And for every such $\,n\,,$ equation (\ref{omega})
determines the particular one-form $\,\omega\,.$ Then, equation
(\ref{omega2}) defines generically the four thermodynamic
variables $\,s\,,$ $\,c\,,$ $\,T\,,$ $\,\mu\,,$ as functions of
the three ones $\,\rho\,,$ $\,p\,,$ $\,n\,;$  the functions
$\,s\,$ and $\,c\,$ result thus involved by one first order
differential equation, namely
\begin{equation} \label{sc}
    c'_p s'_n - c'_n s'_p = - \frac{\rho + p}{n}(c'_p s'_\rho - c'_\rho
s'_p) \,\,\, ,
\end{equation}
which always admits solutions in one of the  unknowns for every
arbitrary choice of the other, such solutions depending (for
example {\em via} an initial problem) on an arbitrary function of
two of the three variables. Then the variables $\,T\,$ and
$\,\mu\,$ are univocally given by
\begin{equation} \label{Tmu}
T = \frac{c'_p}{h}  \ \ , \qquad \mu = - \frac{s'_p}{h}
\end{equation}
where
\begin{equation} \label{h}
h \equiv n(s'_\rho c'_p - s'_p c'_\rho) \ \ \ .
\end{equation}
\vspace{2mm}

\begin{proposition}
The different Z\'arate and Quevedo's binary thermodynamic schemes
that any given perfect fluid solution  $\,(u,\rho,p)\,$ to
Einstein equations admits, are obtained by the free choice of a
$u$-invariant function (determination of the total particle number
density $\,n\,),$ of a function of three variables (say, the
entropy per particle, $\,s(\rho,p,n)\,)$ and of a function of two
variables (say, the fractional concentration at a given value
$\,k\,$  of $\,n\,,$ $\,c(\rho,p,k)\,).$
\end{proposition}

The main statement in \cite{zq} asserts that {\em among the
Szekeres and Stephani families of perfect fluid cosmological
models, the only ones that admit a {\em [Z\'arate and Quevedo's]}
binary thermodynamic scheme are the Friedmann-Robertson-Walker
models.} Proposition 1 shows that this statement is wrong.
Proposition 2 gives an indication of the 'distance' between this
statement and the correct one.

As already said, no complete constraints are known on the space of
formal constitutive relations insuring the physical character of a
model; nevertheless, the richness of the choice of Z\'arate and
Quevedo's schemes stated in Proposition 2 locally guarantees the
usual thermodynamic inequalities (such as $T>0$ or $ 0\leq c \leq
1$).

\section{Some  Z\'arate and Quevedo's thermodynamic schemes for
Szekeres and Stephani universes}

The Szekeres and Stephani universes admitting a standard
thermodynamic scheme, i.e. that of a one-component perfect fluid,
have been considered by different authors.  Bona and Coll
\cite{bc} have shown that the Stephani universes admitting a
thermodynamic scheme are space-times with a 3--dimensional group
of isometries acting on 2--dimensional orbits. This result has
been recovered in \cite{kqs} where the authors have also studied
the thermodynamic Szekeres-Szafron models, and they have shown
that a family of thermodynamic solutions of class II without
symmetries exists.  A different approach has been used in
\cite{cfERE1} in order to study the Szekeres-Szafron space-times
of class II: the solutions which represent a perfect fluid in
local thermal equilibrium have been obtained and the associated
thermodynamic schemes explicitly given. More specific
thermodynamic analysis have been considered in other works. Thus,
Sussman \cite{sus} has presented a family of Stephani universes
which admit, up to a good approximation, an interpretation as
classical mono-atomic ideal gases or as matter-radiation mixtures.
On the other hand, we have obtained all the Stephani universes
which represent a generic ideal gas in local thermal equilibrium
\cite{cfERE2}.

It is worth pointing out that, although in some cases the standard
thermodynamic scheme imposes symmetries on the metrics, these
papers show that {\it  inhomogeneous Szekeres and Stephani
universes exist and are known, that admit a standard thermodynamic
scheme.}

On the other hand, Propositions 1 and 2 show that {\em all}
perfect fluid space-times admit Z\'arate and Quevedo's
thermodynamic schemes, but they give no explicit solutions to
them. The difficulties to find such  explicit solutions associated
with a general perfect fluid lie only in the solution of the two
differential equations (\ref{conN}) and (\ref{sc}) because, as we
have seen, then the entropy per particle $\,s\,$ (or equivalently
the fractional concentration $\,c\,)$ may be chosen arbitrarily,
and the temperature $\,T\,$ and the mixture chemical potential
$\,\mu\,$ are explicitly given by (\ref{Tmu}) and (\ref{h}).

Nevertheless, it is not difficult to find explicit Z\'arate and
Quevedo's thermodynamic schemes for any perfect fluid admitting a
standard thermodynamic one. Remember that a standard thermodynamic
scheme is a perfect fluid solution $(u,\rho,p)$ to equation
(\ref{conT}) that admits a solution $\,n\,$ to equation
(\ref{conN}) and for which there exist functions $\,\bar s\,$ and
$\,T\,$ of, say, the variables $\,\rho\,$ and $\,n\,,$ verifying
\begin{equation} \label{gibbs2}
    T \di \bar{s} = \di(\rho/n) + p \di(1/n) \,\,\, .
\end{equation}

Let $\,(u,\rho,p,n,\bar s,T)\,$ be such a standard thermodynamic
scheme. Choose a space-time function $\,c=c(x^{\alpha})\,$
satisfying $\,0 \leq c(x^{\alpha}) \leq 1\,,$ but otherwise
arbitrary, let $\Phi(c)$ be an arbitrary real function of $\,c\,,$
and define a new function  $\,s \,$ of   $\,\rho\,,$ $\,n\,$ and
$\,c\,$ by
\begin{equation}\label{s}
    s = s(\rho,n,c) \equiv \bar{s}(\rho,n) + \Phi(c) \, ,
\end{equation}
Then, on account of (\ref{gibbs2}) one has:
\begin{equation} \label{gibb3}
    T \di s = \di(\rho/n) + p \di(1/n) + T \Phi'(c) \di c
\end{equation}
and, consequently, calling
\begin{equation} \label{mu}
    \mu = - T \Phi'(c) \,\,\,,
\end{equation}
the set of variables $\,(u,\rho,p,n,s,T,c,\mu)\,$ defines a
Z\'arate and Quevedo's thermodynamic scheme.

Consider then the Szekeres and Stephani universes that admit a
 standard thermodynamic scheme and that are explicitly given
in the above mentioned references [3-7]. On every one of them,
every choice of a pair of functions $\,c(x^\alpha)\,$ and
$\,\Phi(c)\,$ directly defines, by means of equations (\ref{s})
and (\ref{mu}), a Z\'arate and Quevedo's thermodynamic scheme,
i.e. an explicit counterexample of the main statement of
\cite{zq}.

\section {Remarks}

Perhaps it is worthwhile to indicate a pair of incorrect arguments
in Z\'arate and Quevedo's work \cite{zq} that could explain their
erroneous conclusion.

In dealing with Szekeres universes, they select, by integration of
a space-like equation, the sole solution $\,\mu = (\rho + p)/n\,,$
neglecting an arbitrary function of time (see their equation (27)
in \cite{zq}). They neglect it under the argument that $\,\mu\,$
being a function of three thermodynamic variables, and the
expression $\,\mu = (\rho + p)/n\,$ already containing them, one
cannot add to $\,\mu\,$ an arbitrary function of time without
increasing the number of independent variables. This argument is
incorrect because it mix up considerations on space-time variables
and thermodynamic ones without taking care of their specific
connection: the pressure being a function of the sole time in
Szekeres space-times, to add a function of time is nothing but to
add a function of the pressure $\,p\,,$ an addition which is
perfectly admissible in their situation  but that invalidates
their particular expression of Gibbs equation (their equation (28)
in \cite{zq}) and their consequences.

In dealing with Stephani universes, they impose to them two
equations of state, for $\,\mu\,$ and $\,s\,$  (their equation
(40) in \cite{zq}), obtained in the Szekeres case under {\em i)}
the abusive restriction above mentioned and {\em ii)} the
hypothesis of  spatially homogeneous pressure, $\,p,_i = 0\,,$
identically satisfied in Szekeres universes but generically
inadmissible in Stephani ones.

\ack This work has been supported by the Spanish
Ministerio de Ciencia y Tecnolog\'{\i}a, project
AYA2003-08739-C02-02.

\end{document}